\documentclass[prb,aps,amssymb,nofootinbib,floatfix,twocolumn,eqsecnum,showpacs]{revtex4}
\usepackage{graphicx,amsmath}
\newcommand{\be}{\begin{equation}}
\newcommand{\ee}{\end{equation}}
\newcommand{\bea}{\begin{eqnarray}}
\newcommand{\eea}{\end{eqnarray}}
\newcommand{\GG}{\mathcal{G}}

\begin{document}
\preprint{First draft}

\title{Fermi Liquid Theory for the Persistent Current 
Past a Side-Coupled Quantum Dot}

\author{Ian Affleck}
\affiliation{Department of Physics and Astronomy, University of British Columbia, Vancouver, British Columbia, Canada V6T 1Z1}

\author{Erik~S.~S\o rensen}%
\affiliation{Department of Physics and Astronomy, McMaster University, Hamilton,
ON, L8S 4M1 Canada }

\date{\today}

\begin{abstract}
A Fermi Liquid theory is developed for the persistent current past a side coupled
quantum dot yielding analytical predictions for the behavior of the first two
harmonics of the persistent current as a function of applied magnetic flux.
The quantum dot is assumed weakly coupled to a ring of non-interacting electrons
and thus appropriately described as a Kondo impurity. The theory is valid at weak Kondo couplings in the regime where the
system size, $L$, is much larger than the size of the Kondo screening cloud, $\xi_K$.
The predictions of the Fermi Liquid theory are compared to exact diagonalization results for the persistent
current that lend support to the existence of a regime correctly described by this theory.
The finite temperature conductance, at $T\ll T_K$ is also calculated using Fermi liquid theory allowing
the definition of a ``Wilson ratio" relating the conductance and the persistent current.
\end{abstract}

\maketitle
\section{Introduction}
The recent experimental observation of the Kondo effect and 
related physical phenomena in 
quantum dots,\cite{Goldhaber,GG,Cronenwett,Wiel} quantum corrals~\cite{Manoharan}
molecular electronics devices~\cite{Park,Liang}
and carbon nanotubes~\cite{Nygaard} as well as related nano-structures has led to significant renewed
interest in this fundamental effect. For a review of recent progress see Ref.~\onlinecite{KouwenGlazman}.
The Kondo effect in ordinary metals is usually associated with an increase in
the resistance at low temperatures but experiments performed on these
nano-structures~\cite{Goldhaber,Nygaard,Manoharan,Wiel} have shown that the Kondo effect
can also lead to an increase in the conductance that in some settings can reach the unitary limit ($G=2e^2/h$),
completely overcoming the Coulomb blockade. The fundamental and technological interest in the complete
understanding of this phenomenon is therefore considerable. The experiments are typically performed
on semi-conductor quantum dots  where the electron
occupation number on the dot is controllable by a gate voltage. When the occupation
number is odd the quantum dot can act as a spin $s=1/2$ impurity screened by the electrons in
the leads. The screening of the impurity spin by the conduction electrons is, in some circumstances, 
associated with the formation of a ``screening cloud" of size $\xi_K=v_F/T_K$ surrounding
the impurity.~\cite{Nozieres,Sorensen96,Barzykin96,Barzykin98,Cornaglia,Hallberg,Borda} Here $v_F$ is the Fermi velocity and $T_K$ the Kondo temperature.

We note that the picture of a screening cloud of size 
$v_F/\xi_K$ is valid in a one dimensional system and also 
in an infinite two or three dimensional 
system where only one channel (such as the s-wave) couples 
most strongly to the Kondo impurity.  However the behavior 
may be rather different in a disordered 
two or three dimensional box in which all states couple 
strongly to the Kondo impurity.  For the 3-dimensional 
model considered in [\onlinecite{Thimm}], for instance, the effective 
screening cloud is much smaller, $\propto T_K^{-1/3}$. 
See also [\onlinecite{Hand}].

The experimental observation of the screening cloud has proven elusive but recent
experiments~\cite{Wiel} have suggested that it might be observable in finite-size
properties of the conductance when the dot is attached to a single quantum wire
that forms an Aharonov-Bohm  ring of length $L$. A natural extension of these experiments~\cite{Wiel} is to consider
not only the conductance but also finite-size effects observable in the persistent current generated in the 
presence of a magnetic flux, $\Phi$.

The persistent current in quantum dot systems operated in the Kondo limit
has been the topic of a series of recent theoretical 
studies.\cite{
  ZvyaginJLT,
  ZvyaginSchlottmann,
  Ferrari,
  Kang,
  Cho,
  Eckle,
  SAprl,
  SAprb, 
  Zvyaginprl,
  Hu,
  Aligia,
prl.persistent}
Usually, two fundamentally different geometries are considered;\cite{SAprb} the embedded quantum dot (EQD)
and the side-coupled quantum dot (SCQD) and it has been clearly established that the persistent current
has a strong dependence on the parity of the total number of electrons, $N$, in the system.
Here we shall exclusively be concerned with the SCQD.
Suppressing electron spin indices, the Hamiltonian is in this case:
\begin{equation}
H_{\rm SCQD}=-t\sum_{j=0}^{L-1}\left(c^{\dagger}_jc_{j+1}+{\rm h.c.}\right)+H_K,\ \  (c_L\equiv c_0)\label{eq:H}
\end{equation}
where $H_K=J_K{\vec S}\cdot c_0^{\dagger}
\frac{\vec\sigma}{2}c_0$, and
$c_{j\sigma}$ is the electron annihilation operator at site $j$ for spin $\sigma$
and $S^a$ are S=1/2 spin operators. In the following we set $t=1$. We always include the impurity spin as an electron in 
our definition of $N$ but do not count the impurity site in our definition of $L$.  Hence, at half-filling we have $N=L+1$.
The dimensionless magnetic flux, $\alpha\equiv e\Phi/c$ is introduced by appropriately changing the phases
of the hopping terms at the corresponding sites. (We set $\hbar =1$.) 
The persistent current is given
by $j=-edE_0/d\alpha$ where $E_0$ is the ground-state energy. Despite the simplicity
of this geometry, a complete understanding of the behavior of the persistent current
both at half-filling and off half-filling has proved exceedingly difficult to achieve.
In previous work~\cite{SAprb,SAprl} by one of us and P. Simon it was predicted that,
$jL/(ev_F)\to 0$ when $\xi_K/L\to 0$ for both parities of $N$ in contrast to the results of Refs~\onlinecite{Eckle,Cho}.
However, subsequent work~\cite{prl.persistent} confirmed this prediction at half-filling but showed 
both analytical and numerical evidence for additional terms off half-filling for $N$ even.

In the present paper we develop a Fermi liquid theory (FLT) for the persistent current
past a SCQD, valid for $1\ll \xi_K\ll L$. The theory provides independent
confirmation of the above mentioned prediction at half-filling~\cite{SAprb,SAprl} as
well as additional analytical evidence for the additional terms off half-filling for $N$ even.
The theory yields detailed predictions about the finite-size $L$-dependence of the
first two harmonics of the persistent current, using a single parameter, $1/T_K$. 
Section~\ref{sec:flt} outlines the FLT in considerable detail while section~\ref{sec:half}
presents our numerical results at half-filling and section~\ref{sec:offhalf} our numerical
results off half-filling.  In Sec.~\ref{sec:conductance}  we calculate the finite 
temperature conductance past the quantum dot, at $T\ll T_K$. 
This enables us to calculate a generalized ``Wilson ratio'' 
relating the persistent current for a finite ring at $T=0$ 
with the conductance at finite $T$ for the same quantum dot
side-coupled to infinite leads. This could be useful 
in an experimental context since a prior infinite leads 
measurement of the low $T$ conductance would fix 
the precise numerical coefficient in the finite size 
dependence of the persistent current at $T=0$.  

\section{Fermi Liquid Theory for the Persistent Current\label{sec:flt}}
Kondo physics at low energy scales and long length scales, 
$T\ll T_K$, $L\gg \xi_K$, 
can be described by an effective Hamiltonian not containing the 
impurity spin, since it is screened by the conduction electrons 
and the energy scale associated with breaking this spin singlet 
is of $O(T_K)$. The low energy quasi-particles propogate freely 
except that they obey a modified boundary condition which 
can be thought of as arising because  
they must stay in wave-functions orthogonal to that  
of the screening electron. This modified boundary condition 
is equivalent to a $\pi /2$ phase shift for even-channel 
electrons (or  for s-wave electrons 
in the three dimensional, spherically symmetric version 
of the model).  In addition to the 
modified boundary condition an interaction term is generated 
in this effective Hamiltonian which is only non-vanishing 
near the impurity location, at $|x|<\xi_K$.  In a long 
wavelength effective Hamiltonian this interaction appears right at 
the origin.  This interaction is
 irrelevant  in the renormalization 
group sense; treating it in perturbation theory, its effects 
become weaker and weaker at lower energies and longer lengths. 
Building on earlier work by Wilson,~\cite{wilson} Nozi\`eres~\cite{Nozieres}
developed an elegant description of this physics, following 
the Landau approach to Fermi liquid theory, by writing 
an expression for the phase shift which depends on the 
energy of the electron being scattered and also on the 
local density of electons at the origin. In [\onlinecite{AL}]
this approach was transcribed into an effective Hamiltonian, 
an approach which is simpler to deal with in some situations 
and more in accord with modern renormalization group (RG)
 techniques.  (In a similar 
way, the Landau approach to Fermi liquids can be recast in 
terms of an effective Hamiltonian which can be studied 
by RG methods.  See, for example, [\onlinecite{shankar}].)  Furthermore, 
in [\onlinecite{AL}] it was shown how this 
effective Hamiltonian could be uniquely determined, up to one 
overall coupling constant with dimensions of inverse energy, $1/T_K$, 
 by using the one-dimensional (1D) formulation of 
the Kondo model and 1D spin-charge separation. Essentially {\it all} low 
energy long distance properties of the Kondo model can be 
calculated by perturbation theory in this Fermi liquid interaction. 
They are generally proportional to the first or second power 
of the coupling constant, $1/T_K$ or $1/T_K^2$. It is not an 
easy matter to determined the precise value of $T_K$ for 
a given microscopic model. However, for a weak bare Kondo 
coupling, $\nu J\ll 1$ (where $\nu$ is the density of states) it
is known that $T_K$ is exponentially small, $T_K\propto D\exp [-1/(J\nu )]$,
where $D$ is a cut-off or bandwidth scale. Furthermore, when the 
bare coupling is weak, the physics is universal at all energy 
scales $\ll D$.  This implies, in particular, that 
the ratios of various physical quantities calculated in 
perturbation theory in $1/T_K$ are universal.  The most famous 
such universal ratio, between the impurity susceptibility 
and specific heat, is known as the Wilson ratio. However, many 
more such universal ratios can be readily calculated. In this 
section we extend extend this approach to the persistent current. 

\subsection{The Boundary Conditions}
Since Fermi liquid theory applies at low energies and long distances, 
it is convenient to   
linearize the spectrum around
the Fermi points $\pm k_F$ and to introduce left (L) and right (R) moving
chiral fields:
\begin{equation}
c_x\sim e^{ik_F x}\psi_R(x)+e^{-ik_Fx}\psi_L(x).
\label{eq:2.10}
\end{equation}
The non-interacting part of the continuum Hamiltonian becomes:
\begin{equation} H_0=\int_{0}^L\left[ \psi_R^{\dagger}i\partial_x\psi_R
-\psi_L^{\dagger}i\partial_x\psi_L\right] .\end{equation}
We have set $v_F=1$. 
The Kondo interaction becomes:
\begin{equation} H_K\to J_K\vec S\cdot [\psi^\dagger_R(0)+\psi^\dagger_L(0)]
{\vec \sigma \over 2} [\psi_R(0)+\psi_L(0)].\end{equation}
The strong coupling fixed point of the Kondo problem corresponds to 
the boundary condition:
\begin{equation} c_0\propto \psi_L(0)+\psi_R(0)=0.\label{bc0}\end{equation}
We may think of the sceening electron as sitting at the origin, $j=0$ 
and the other electrons (low energy quasi-particles) must not 
enter or leave the origin in order to avoid breaking up the Kondo singlet. 
Considering the model defined on a ring, and ignoring, for the moment,
the magnetic flux, there is a second boundary condition at the 
strong coupling fixed point:
\begin{equation} c_L\propto e^{ik_F L}\psi_R(L)+e^{-ik_FL}\psi_R(L)=0.\label{bcL}\end{equation}
This strong coupling fixed point corresponds to an open chain 
of $L-1$ sites [$j=1,2,3,\ldots (L-1)$] with no impurity spin. 
Initially we consider half-filling. 
In this case, for such an open chain, $k_F=\pi /2$ 
regardless of the parity of $L$. Since the boundary conditions 
of Eqs. (\ref{bc0} ,\ref{bcL}) are true at all times, while 
$\psi_{R/L}$ is a function of $t\mp x$ only, it follows from Eq. (\ref{bc0}) that 
we can regard $\psi_R(x)$, for $x>0$ as the analytic continuation of $\psi_L(x)$ 
to the negative $x$ axis:
\begin{equation} \psi_R(x)=-\psi_L(-x),\ \  (x>0).\label{ac1}\end{equation}
Likewise, Eq. (\ref{bcL}) implies that we can also regard 
$\psi_R(x)$ as the analytic continuation of $\psi_L(x)$:
\begin{equation} \psi_R(L-x)=-e^{-2ik_FL}\psi_L(L+x).\ \  (x>0)\label{ac2}\end{equation}
Letting $x\to L$ in Eqs. (\ref{ac2}), we see that:
\begin{equation} \psi_L(2L)=e^{2ik_FL}\psi_L(0).\end{equation}
At half-filling this becomes:
\begin{equation} \psi_L(2L)=(-1)^{L}\psi_L(0),\label{pbc}\end{equation}
periodic for $L$ even and anti-periodic for $L$ odd. We 
may formulate the model in terms of left-movers only with 
these periodic or anti-periodic boundary conditions. 

\subsection{The Fermi Liquid Interaction with Zero Flux}
We now wish to write the Fermi liquid interaction in terms of 
this left-moving formulation of the model. This must involve 
the even channel fermions (even under $x\to L-x$) only since only the 
even channel appears in the Kondo interaction. Furthermore, 
it only involves this channel near $x=0$.  Consider for example 
the even channel fermions in the lattice model at a distance 
of 1 site from the origin:
\begin{widetext}
\begin{equation} c_1+c_L \propto e^{ik_F}\psi_R(1)+e^{-ik_F}\psi_L(0)
+e^{ik_F(L-1)}\psi_R(L-1)+e^{-ik_F(L-1)}\psi_L(L-1).\end{equation}
\end{widetext}
We now use the facts that $k_F=\pi /2$, the continuum fields $\psi_L(x)$ 
and $\psi_R(x)$ are slowly varying on the lattice scale and 
the boundary conditions of Eq. (\ref{bc0}, \ref{bcL}) to write this 
in the purely left-moving formulation as:
\begin{equation} c_1+c_L\propto e^{-i\pi /2}\psi_L(0)+e^{-i\pi (L-1)/2}\psi_L(L).\label{psie}\end{equation}
We expect {\it only} this combination of fields to appear 
in the Fermi liquid interaction, up to higher dimension operators 
involving derivatives of the fermion fields. This follows from 
observing, using the boundary conditions of 
Eq. (\ref{bc0}, \ref{bcL}) and the 
fact that the continuum fields vary slowly, that {\it any} non-vanishing even lattice fermion field 
near the origin is proportional to this one.
For $j$ {\it even} and $j\ll \xi_K$ one finds:
\begin{equation} 
c_j+c_{L-j}\approx 0.
\end{equation} 
While for $j$ {\it odd} and $j\ll \xi_K$ we have:
\begin{eqnarray} 
\lefteqn{c_j+c_{L-j}\propto e^{-i\pi j/2}\psi_L(0)+e^{-i\pi (L-j)/2}\psi_L(L)}&&\nonumber\\
&&=(-1)^{(j-1)/2}\left[e^{-i\pi /2}\psi_L(0)+e^{-i\pi (L-1)/2}\psi_L(L) \right],\nonumber\\
\end{eqnarray}
which is proportional to Eq.~(\ref{psie}).

Once we have written the Kondo interaction in terms of 
the even sector fermions only, it is convenient 
to bosonize, introducing separate spin and charge bosons. 
Importantly, the Kondo interaction involves 
only the spin bosons in the even sector. 
It 
then follows that the Fermi liquid interactions 
can be written in terms of these bosons only. The only possible 
dimension 2 operator, in the spin sector, which respects the SU(2) 
symmetry is $\vec J_e^2(0)$, the square of the spin
density operator. Going back 
to the fermion representation:
\begin{equation}
\vec J_e(0)\equiv \psi^\dagger_e(0){\vec \sigma \over 2}
\psi_e(0).
\end{equation}
where, from Eq. (\ref{psie}), 
\begin{equation} \psi_e(0)\propto {\psi_L(0)-e^{-i\pi L/2}\psi_L(L)\over \sqrt{2}}.\end{equation}
Thus we write the leading irrelevant Fermi liquid interaction, 
for the case, $L$ even as:
\begin{eqnarray}
H_{int}&=&{-8\pi \over 3T_K}\vec J_e^2(0)\nonumber\\
&=&{-2\pi \over 3T_K}
\{[\psi_L(0)-e^{-i\pi L/2} \psi_L(L)]^\dagger {\vec \sigma \over 2}\nonumber\\
&\times&
[\psi_L(0)-e^{-i\pi L/2}\psi_L(L)]\}^2.
\label{Hint}
\end{eqnarray}
We have written the coupling constant in front of this operator 
as $8\pi /3T_K$. It follows from standard scaling arguments 
that this coupling constant should be of order the RG cross-over 
scale.  In general, it should be written as $1/T_K$ times 
a dimensionless constant of O(1).  All low energy propertites 
of the system can be determined from this interaction term, 
including the impurity susceptibility and specific heat, 
for example.  The sign in Eq. (\ref{Hint}) is known 
to be the correct one since it gives the correct (positive) 
sign for these two quantities. The factor of $8\pi/3$ is 
purely a matter of a convention, or a precise definition 
of what is meant by $T_K$.  Unfortunately, a great number of 
different conventions for $T_K$ are in current use. We have 
chosen here to use the same convention as used by 
Nozi\`eres~\cite{Nozieres} (who also 
referred to $1/T_K$ as $\alpha$) and by Glazman and Pustilnik~\cite{GP}.
In Appendix B we briefly review other definitions of $T_K$ in popular 
use and the constant factors relating them to each other. 
 
Strictly speaking, one other operator of dimension 2 is permitted, 
$J_e(0)^2$, the square of the charge current, rather 
than the spin current.  As mentioned above, naively the charge sector
decouples from the Kondo interactions so that this operator 
should apparently not be generated.  However, irrelevant 
operators (at the weak coupling fixed point) couple 
charge and spin sectors together ultimately permitting this 
operator to occur. However, the coefficient of this 
operator is expected to be $O(1/D)$, where $D$ is the 
bandwidth ($t$ in the tight-binding model), rather than $O(1/T_K)$. 
In the weak coupling limit, where $T_K\ll D$, this other 
interaction can be ignored. 

\subsection{The Fermi Liquid Interaction with Finite Flux}
So far we haven't mentioned the flux, $\Phi = c\alpha /e$.
 The phases in the hopping terms corresponding 
to any desired flux can be inserted anywhere in the ring. They 
can be moved around freely by phase redefinitions of 
the electron fields (gauge transformations). 
For purposes of understanding the strong coupling 
fixed point it is convenient to imagine that they 
are inserted far from the impurity compared to $\xi_K$ 
so that the strong coupling boundary conditions, discussed 
above, are unaffected by the flux.  Of course, this 
is only possible for $L\gg \xi_K$ but it is precisely 
that limit which we are now considering.  The Fermi 
liquid theory only applies in that limit. It can 
be seen that adding  phases to hopping terms neccessarily 
couples even and odd sectors together, since it 
breaks parity. For this reason, it is more convenient 
to go back to left and right movers and then ultimately 
to left movers only on a ring of length $2L$, as 
discussed above. It is also most convenient to 
put the phase at the origin {\it after} establishing 
the strong coupling b.c. This amounts to:
\begin{equation}
\psi_L(L)\to e^{-i\alpha}\psi_L(L).
\end{equation}
Hence the Ferm liquid interaction becomes:
\begin{eqnarray}
H_{int}
&=&{-2\pi \over 3T_K}
\{ [\psi_L(0)-e^{-i(\pi L/2+\alpha )} \psi_L(L)]^\dagger 
{\vec \sigma \over 2}\nonumber\\
&\times&
[\psi_L(0)-e^{-i(\pi L/2+\alpha )} \psi_L(L)]\}^2.
\label{flt}
\end{eqnarray}

\subsection{Perturbation Theory}
The remaining calculations are quite straightforward. 
We simply do first order perturbation theory in $1/T_K$ 
for the groundstate energy, imposing periodic 
or anti-periodic b.c.'s, Eq. (\ref{pbc}),  
on the purely left-moving fermion fields. 

Now we turn to calculating the expectation value of 
$H_{int}$.  From Eq.~(\ref{flt}), recalling that 
$L=N-1$ at half-filling, it is clear that it is advantageous to define
a shifted flux $\tilde\alpha$ in the following manner:
\begin{eqnarray}
\tilde\alpha &=& \alpha+\pi(N-1)/2 \ \ {\rm (N\ odd)}\nonumber\\
\tilde\alpha &=& \alpha+\pi N/2 \ \ {\rm (N\ even)}.
\label{eq:alpha}
\end{eqnarray}
[This definition can been seen largely as a way of defining the coefficient in
front of $\sin(\tilde\alpha)$ in the persistent current to be positive.]
Using this definition,
we first rewrite $H_{int}$ as:
\begin{widetext}
\begin{eqnarray}
H_{int}&=&{ \pi \over 6T_K}\bigl\{ e^{2i\tilde\alpha}[\psi^\dagger (L)\vec \sigma \psi (0)]^2
-2ie^{i\tilde \alpha} \psi^\dagger (L) \vec \sigma \psi (0)
\cdot\left[\psi^\dagger (0) \vec \sigma \psi (0)
+\psi^\dagger (L) \vec \sigma \psi (L)\right]
+h.c. + \ldots \bigr\}, \ \  {\rm (N\ even)}\nonumber \\
H_{int}&=&{ \pi \over 6T_K}\bigl\{- e^{2i\tilde\alpha}[\psi^\dagger (L)\vec \sigma \psi (0)]^2
+2e^{i\tilde \alpha} \psi^\dagger (L) \vec \sigma \psi (0)
\cdot\left[\psi^\dagger (0) \vec \sigma \psi (0)
+\psi^\dagger (L) \vec \sigma \psi (L)\right]
+h.c. + \ldots \bigr\}. \ \  {\rm (N\ odd)}\nonumber\\
\label{hint2}
\end{eqnarray}
\end{widetext}
Here we have dropped the $L$ subscripts and also dropped terms 
which are independent of $\alpha$ and hence won't contribute 
to the current in first order in $1/T_K$. 
The first term can be rewritten:
\begin{equation}
 H_{int}^{(1)}={\pi (-1)^N\over T_K}e^{2i\tilde\alpha}\psi^{\uparrow \dagger}(L)
\psi_{\uparrow}(0)\psi^{\downarrow \dagger}(L)
\psi_{\downarrow}(0)+ h.c.
\end{equation}
We now evaluate 
the various terms using Wick's theorem. For this we need the 
equal time propogator for a left mover with anti-periodic and periodic boundary conditions 
on an interval of length $2L$.  

\subsubsection{$N$ Even at Half-filling}
For the case $N$ even, with anti-periodic 
boundary conditions, this is:
\begin{eqnarray}
 <\psi^{\dagger \alpha}(x) \psi_\beta (0)>&=&\delta^\alpha_\beta 
{1\over 2L}\sum_0^\infty e^{-i\pi (m+1/2)x/L}\nonumber\\
&=&{-i
\delta^\alpha_\beta \over 4L\sin (\pi x/2L)}.
\label{prop}
\end{eqnarray}
In particular:
\begin{equation}
<\psi^{\dagger \alpha}(L) \psi_\beta (0)>={-i
\delta^\alpha_\beta \over 4L}.\label{prope}
\end{equation}
Of course, 
\begin{equation}
<\psi^{\dagger \alpha}(L) \psi_\beta (0)>=
-<\psi_\beta (0) \psi^{\dagger \alpha} (L)>.\label{ob}
\end{equation}
We also need an expression for $<\psi^{\dagger\alpha} (0)\psi_\beta (0)>$.
The sum in Eq. (\ref{prop}) is ultraviolet divergent so we introduce 
a (dimensionless) cut-off, an upper bound on the summation variable 
$m$: $m < D$, giving:
\begin{equation}
<\psi^{\dagger\alpha} (0)\psi_\beta (0)>={\delta^{\alpha}_\beta D\over 2L}.
\label{D} 
\end{equation}
We will see that our results for the current do not depend 
on the value of $D$. It follows from PH symmetry that:
\begin{equation}
<\psi_\beta (0) \psi^{\dagger\alpha} (0)>={\delta^\alpha_\beta
D\over 2L}.\label{PH}
\end{equation}
Furthermore 
\begin{equation}
<\psi_\beta (0) \psi^{\dagger\alpha}(0)>=
<\psi_\beta (L) \psi^{\dagger\alpha}(L)>, \label{0L}
\end{equation}
as follows from the reflection symmetry of the  model which 
takes $x\to L-x$. 

The $\cos 2\alpha$ term is:
\begin{equation}
E_0^{(2)}={-\pi \over T_K}e^{2i\tilde\alpha}\left( {-i\over 4L}\right)^2+h.c. 
={\pi \over 8T_KL^2}\cos 2\tilde\alpha ,
\end{equation}
With a current:
\begin{equation}
j^{(2)}={-edE_0/d\tilde\alpha}={e\pi \over 4T_KL^2}\sin 2\tilde\alpha\ (N\ {\rm even}).\label{2}
\end{equation}

Now consider the $\cos \tilde\alpha$ term in the energy. This is actually zero 
for $N$ even but let's go through some steps which are 
useful also for $N$ odd. We use Wick's theorem to write this as:
\begin{eqnarray} 
&&<\psi^\dagger (L)\vec \sigma \psi (0)\cdot 
[\psi^\dagger (0)\vec \sigma \psi (0)+\psi^\dagger (L)\vec \sigma \psi (L)]>\nonumber\\
&&=2<\psi^\dagger (L)\vec \sigma \psi (0)>\cdot 
<\psi^\dagger (0)\vec \sigma \psi (0)>\nonumber \\
&&+<\psi^{\alpha \dagger}(L)\psi_{\delta}(0)><\psi_\beta (0)
\psi^{\gamma \dagger}(0)>\vec \sigma^\beta_\alpha \cdot 
\vec \sigma^{\delta}_\gamma \nonumber\\
&&+ <\psi^{\alpha \dagger}(L)\psi_{\delta}(L)><\psi_\beta (0)
\psi^{\gamma \dagger}(L)>\vec \sigma^\beta_\alpha \cdot 
\vec \sigma^{\delta}_\gamma. \nonumber\\
\label{wick} 
\end{eqnarray}
For $N$ even it is easy to see that the first term in Eq. (\ref{wick}) 
vanishes and the last two cancel using Eqs. (\ref{prope})-(\ref{0L}).

\subsubsection{$N$ Odd at Half-filling}
Now consider $N$ odd. The strong coupling groundstate has 
spin -1/2; we choose the state with total $S^z=1/2$.  The 
propogators are now different for spin up or down.  We find:
\begin{equation}
<\psi^{\uparrow \dagger}(x)\psi_{\uparrow} (0)>=
{1\over 2L}\sum_{m=0}^\infty e^{-i\pi mx/L}={-i\over 4L}
{e^{i\pi x/2L}\over \sin (\pi x/2L)}.
\end{equation}
On the other hand, the $\downarrow$ propagator is the 
same except that the $m=0$ term is omitted from the sum, 
subtracting $1/2L$ to the final result. Thus:
\begin{equation}
<\psi^{\alpha \dagger}(L)\psi_\beta(0)>={(\sigma^z)^\alpha_\beta
\over 4L}.\label{propodd}
\end{equation}
 Again we impose a cut-off so:
\begin{equation}
<\psi^{\uparrow \dagger}(0)\psi_\uparrow (0)>={D\over 2L}.
\end{equation}
Now we have:
\begin{equation}
<\psi_\uparrow (0) \psi^{\uparrow \dagger}(0)>={D-1\over 2L}.\label{PHoup}
\end{equation}
The $(-1)$ reflects the breaking of particle-hole (P-H) symmetry. 
We have one unpaired spin up electron, sitting right 
at the Fermi surface.  Furthermore,
\begin{equation}
<\psi^{\downarrow \dagger} (0) \psi_{\downarrow }(0)>={D-1\over 2L}
\end{equation}
and
\begin{equation}
<\psi_\downarrow (0) \psi^{\downarrow \dagger}(0)>={D\over 2L}.\label{PHod}
\end{equation}
We may combine these as:
\begin{eqnarray}
<\psi^{\alpha \dagger} (0)\psi_\beta (0)>&=&{D-1/2\over 2L}
\delta^{\alpha}_{\beta} + {1\over 4L}(\sigma^z)^{\alpha}_{\beta} 
\nonumber \\
<\psi_\alpha (0)\psi^{\beta \dagger} (0)>&=&{D-1/2\over 2L}
\delta_{\alpha}^{\beta} - {1\over 4L}(\sigma^z)^{\alpha}_{\beta}.
\end{eqnarray}
We now substitute this into Eq. (\ref{wick}).  It is 
easy to see that the terms containing $D$ all cancel, for 
the same reasons that the entire expression vanishes for $N$ even. 
The remaining terms are:
\begin{eqnarray}
&& <\psi^\dagger (L)\vec \sigma \psi (0)\cdot 
[\psi^\dagger (0)\vec \sigma \psi (0)+\psi^\dagger (L)\vec \sigma \psi (L)]>\nonumber\\
&&=\left({1\over 4L}\right)^2[2tr (\vec \sigma \sigma^z) \cdot 
tr (\vec \sigma \sigma^z)-2tr (\vec \sigma \sigma^z\cdot \vec \sigma \sigma^z)]\nonumber\\
&&= \left({1\over 4L}\right)^2[8+4]\nonumber\\
&&={3\over 4L^2}.
\end{eqnarray}
Thus the $\cos \tilde\alpha$ term in the energy, for $N$ odd, is:
\begin{equation}
E^{(1)}_0={\pi \over 6T_K}2e^{i\tilde\alpha}{3\over 4L^2}+c.c.={\pi \over 2T_KL^2}
\cos \tilde \alpha .
\end{equation}
The term in the current is:
\begin{equation}
j^{(1)}={e\pi \over 2T_KL^2}\sin \tilde \alpha\ (N\ {\rm odd}).\label{eq:sin}
\end{equation}

The $\sin (2\tilde\alpha)$ in the current for odd $N$ is given again by 
Eq. (\ref{2}), with the opposite sign: 
\begin{equation}
j^{(2)}=-{e\pi \over 4T_KL^2}\sin 2\tilde\alpha\ (N\ {\rm odd}).
\end{equation}
Note that the minus 
sign which follows from spin up and down Green's functions having 
opposite sign for $N$ odd, Eq. (\ref{propodd}),   replaces 
the minus sign due to the factors of $i$ in the Green's function 
for $N$ even, Eq. (\ref{prope}), leaving only the minus 
sign from Eq. (\ref{hint2}).

We have so far set $v_F=1$.  We summarize our results on the persistent current, for $\xi_K\ll L$
below, reinserting a factor of $v_F^2$ by dimensional analysis and replacing 
$v_F/T_K$ by $\xi_K$:
\bea j_e &\to& {ev_F\over L}{\xi_K\over L}{\pi \over 4}\sin 2\tilde \alpha\nonumber \\
j_o &\to & {ev_F\over L}{\xi_K\over L}\left[ {\pi \over 2}\sin \tilde \alpha -{\pi \over 4}\sin 2\tilde \alpha 
\right].\label{jfinal}\eea
It was argued earlier\cite{SAprl}, that both $j_e$ and $j_o$ can be written as $ev_F/L$ times 
universal scaling functions of $\xi_K/L$ and $\tilde \alpha$. Eq. (\ref{jfinal}) indeed has that form. 

\subsubsection{Off Half-filling}
Now consider the case away from 1/2-filling, where P-H  
symmetry is broken.  
 Here we refer to particle-hole symmetry breaking by 
an amount of $O(1)$ not an amount of $O(1/L)$ 
as occurs, even at 1/2-filling for $N$ even.
Note that we continue to choose the reduced band to be symmetric around $k_F$:
$k_F-\Lambda \leq k \leq k_F+\Lambda$. 
However, the 
process of integrating out wave-vectors further away from $k_F$ is 
not P-H symmetric, when $k_F$ does not have the P-H symmetric value, 
so we expect to generate PH symmetry breaking terms
in $H_{int}$. The most important such term, 
which can lead to a current of $O(1/L)$, is:
\begin{equation}
H_2 = -e^{i\pi (N-1)/2}\lambda e^{i \alpha} \psi^\dagger (L)\psi (0)
+ h.c.
\label{H2}\end{equation}
where $\lambda$ is a real coupling constant. Here
the phase in this expression is determined by parity, i.e. 
by the fact that 
it arises from a term $-2\lambda \psi^\dagger_e\psi_e$ using 
Eq. (\ref{psie}).
This term only occurs when P-H symmetry is broken.  We expect $\lambda \propto (J_K/t)^2$ 
where $t$ is the original bandwidth.  

More generally, we 
might start with a microscopic model which includes 
second neighbor tunnelling past the quantum dot:
\begin{equation}
\delta H = -t_2(c_{L-1}^\dagger c_1+h.c.).\end{equation}
Such a term breaks P-H symmetry even at half-filling 
and leads to a term in the low energy effective Hamiltonian 
of the form of $H_2$ in Eq. (\ref{H2}) with $\lambda \propto t_2$.

If the system is very close to half-filling, $n-1=\delta n\ll 1/\xi_K$, 
and we begin with a pure Kondo model with no potential scattering 
or direct tunnelling, then we can calculate the value of 
$\lambda$, in $H_2$ directly from the Fermi liquid interaction
$H_{int}$ of Eq. (\ref{hint2}).  To do this, it is convenient 
to integrate out wave-vectors in the RG transformation symmetrically 
around $k=\pi /2$ even thought this is no longer $k_F$, which has 
the value:
\begin{equation} k_F=(\pi /2)(1+\delta n).\end{equation}
 In this 
way, $\lambda$ is not generated during the RG transformation, 
and we obtain only the interaction $H_{int}$. However, the 
low energy effective theory inherits an asymmetric cut-off:
\begin{equation} k_F-\Lambda -(\pi /2)\delta n < k < k_F-(\pi /2)\delta n+\Lambda .\end{equation}
We may now generate $H_2$ from $H_{int}$ by simply normal 
ordering $H_{int}$. This follows since Eqs. (\ref{D}, \ref{PH}) 
are now modified to:
\bea <\psi^{\dagger \alpha}(0)\psi_\beta (0)>&=&
\delta^\alpha_\beta \left[ {D\over 2L}+{\delta n\over 4}\right] 
\nonumber \\
<\psi_\beta (0)\psi^{\dagger \alpha}(0)>&=&
\delta^\alpha_\beta \left[ {D\over 2L}-{\delta n\over 4}\right]
\label{noPH}\eea
We now write $H_{int}$ as a normal ordered part plus 
a correction of the form of $H_2$. This follows using:
\begin{widetext}
\bea  &&\psi^\dagger (L)\vec \sigma \psi (0)\cdot \left[ \psi^\dagger (0)\vec \sigma \psi (0)
+\psi^\dagger (L)\vec \sigma \psi (L)\right]
=:\psi^\dagger (L)\vec \sigma \psi (0)\cdot \left[ \psi^\dagger (0)\vec \sigma \psi (0)
+\psi^\dagger (L)\vec \sigma \psi (L)\right] :
\nonumber \\ 
&+&\left[ \psi^{\alpha \dagger}(L)\psi_{\delta}(0)<0|\psi_\beta (0)\psi^{\gamma \dagger}(0)|0>
-\psi^{\gamma \dagger}(L)\psi_\beta (0)<0|\psi^{\alpha \dagger}(L)\psi_\delta (L)\right]
(\vec \sigma^\beta_\alpha \cdot \vec \sigma^\delta_\gamma ).\label{normord}\eea
\end{widetext}
Using Eq. (\ref{noPH}) we see that the two terms in the second line of Eq. (\ref{normord})
don't cancel, away from half-filling, yielding instead:
\begin{widetext}
\begin{equation} \psi^\dagger (L)\vec \sigma \psi (0)\cdot \left[ \psi^\dagger (0)\vec \sigma \psi (0)
+\psi^\dagger (L)\vec \sigma \psi (L)\right]
=:\psi^\dagger (L)\vec \sigma \psi (0)\cdot \left[ \psi^\dagger (0)\vec \sigma \psi (0)
+\psi^\dagger (L)\vec \sigma \psi (L)\right] :-{3\delta n\over 2}\psi^\dagger (L)\psi (0).
\end{equation}
\end{widetext}

Inserting this expression into Eq. (\ref{hint2}) 
gives a quadratic term of the form of $H_2$ with:
\begin{equation} \lambda = {\pi \delta n\over 2T_K},\label{lambdan}\end{equation}
for $N$ even.

We may now evaluate the additional terms in the current 
arising from $H_2$, using perturbation 
theory in $H_{int}$. We only discuss first order perturbation 
theory.  
We evaluate $<H_2>$ using Eq. (\ref{prope}). This gives an
additional current:
\begin{equation}
\delta j = (e\lambda /L)\sin \tilde \alpha ,\label{jPH}
\end{equation}
for $N$ even. For $N$ odd, there 
is no extra contribution to first order in $\lambda$ since 
$<\psi^{\dagger \alpha}(L)\psi_\beta (0)>\propto (\sigma^z)^\alpha_\beta$.
Hence, to first order, we expect that in the limit of small $\delta n$ the current for $N$ odd,
$j_o$, remain unchanged with respect to it's value at half-filling.

Note however, that  we have only evaluated the contributions 
of $H_{int}$ to the current in first order perturbation theory in 
 $H_2$. Thus we haven't ruled out the possibility of 
a term of $O(1/L)$ in the current for $N$ odd, with broken PH symmetry. 

Adding the universal result for the PH symmetric case to this 
correction from PH symmetry breaking (and reinserting the factor of $v_F$ previously set to one) gives a persistent current, 
for $N$ even:
\begin{equation} j_e = {ev_F\over L}\left[{\pi \xi_K\over 4L} \sin (2\tilde \alpha )+\lambda \sin (\tilde \alpha )
\right].\end{equation}
The non-universal term arising from PH symmetry breaking, proportional to $\lambda$  will always dominate at sufficiently 
large $L$.  However, provided that the dimensionless coupling, $\lambda \ll 1$, 
as we expect for small bare Kondo coupling and not too strong direct tunnelling 
across the quantum dot, the first, universal, term will dominate over 
the range of lengths:
\begin{equation} \xi_K\ll L \ll \xi_K/\lambda .\end{equation}

The presence of a term in $j_e$ proportional to $\sin(\tilde\alpha)$ in the absence of PH symmetry
was shown in Ref.~\onlinecite{prl.persistent} using strong coupling perturbation theory. The above derivation
provides further independent analytical evidence for such a term.

\section{Numerical Results\label{sec:numres}}
The Fermi liquid theory results for the persistent current developed in the previous section are parameterized
in terms of $T_K$ and should be valid in the regime $T\ll T_K$, $1\ll \xi_K\ll L$. 
For a useful numerical test of these results it is therefore necessary to study
ratios of the Fourier components of the persistent current that become {\it independent} of $T_K$ and
can be tested at $T=0$ using exact diagonalization (ED) methods. Furthermore, in order for $1\ll \xi_K$ to hold,
$J_K$ should not be too large. A more severe constraint is that
the system sizes should satisfy $L\gg\xi_K$. With an exponentially diverging $\xi_K$ at small $J_K$
we see that we quickly leave the regime of validity of Fermi liquid theory as $J_K\to 0$ for the
rather modest systems sizes we can treat using ED. Clearly, the interesting regime is then intermediate values
for $J_K$.
Our approach is then to calculate the different
Fourier components, $a_n$,  of the persistent current, $j$, as a function of the flux, $\alpha$.
Our convention for the
Fourier components are $j(\tilde\alpha)=\sum_n a_n \sin(n\tilde\alpha)$. 
Note that, a very precise determination of
the $\alpha$ dependence is necessary in order to determine the Fourier components reliably and we typically
use 200 values for $\alpha$.
The numerical results can then be compared to 
the predictions of the Fermi liquid theory as well as results previously obtained using strong
coupling perturbation.\cite{prl.persistent}

\subsection{At Half-filling\label{sec:half}}
We start by discussing our numerical ED results obtained at half-filling
with systems sizes of $L=12-15$. Due to the lack of symmetry in these systems it is difficult to reach
larger sizes and the largest system $(L=14,N=15)$ required a diagonalization of a matrix of size
$20,796,633$ for each value of $\alpha$ after symmetry reductions.
\begin{figure}[t]
\begin{center}
\includegraphics[clip,width=8cm]{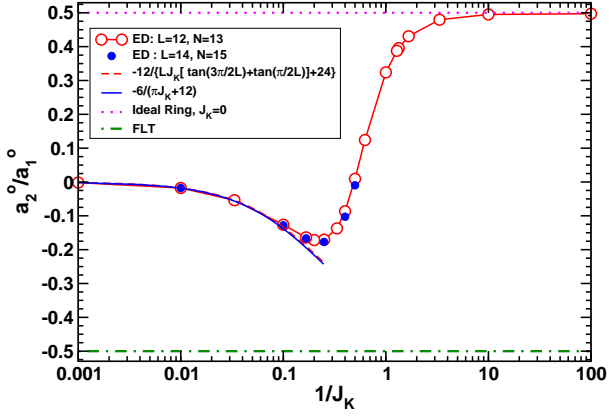}
\caption{The ratio, $a_2^o/a_1^o$, of the two first Fourier coefficients of the current for
the SCQD at 1/2-filling, for $N$ odd. $\circ$ indicates ED results with $L=12$ and $N=13$ while
$\bullet$ indicates ED results with $L=14$ and $N=15$. The dashed line (barely visible) is the  strong coupling
perturbative result, Eq.~(\ref{eq:stronghalf}), for this ratio with $L=12$, the solid line is $L\to\infty$
the strong coupling perturbative result, Eq.~(\ref{eq:thermostronghalf}), the dashed dotted line the fermi liquid 
result, Eq.~(\ref{eq:flthalf}), and the dotted line
the ideal ring result, Eq.~(\ref{eq:zerohalf}).
\label{fig:HalfRatOddOdd}}
\end{center}
\end{figure}

In Ref~\onlinecite{prl.persistent} strong coupling perturbative results were given for the
persistent current at half-filling past the SCQD. It was shown that,
for odd or even $N$, 
\begin{eqnarray}
j_oL/e&\approx &\frac{32}{9J_K^2}[
\tan (\frac{\pi}{2L})+
\tan (\frac{3\pi}{2L})
]\sin \tilde \alpha 
\nonumber \\
&&+\frac{128}{3J_K^3L}[2\sin \tilde \alpha -\sin (2\tilde \alpha )]
\nonumber \\
j_eL/e&\approx &\frac{32}{3J_K^3L}[1+1/\cos (\pi /L)]^2\sin 2\tilde\alpha .
\label{eq:jlarge}
\end{eqnarray}
As elsewhere, $L$ does {\it not} include
the impurity site while $N$ {\it does} include the impurity electron.
The definition of $\tilde\alpha$ is also the same:  For
$N$ odd $\tilde\alpha = \alpha+\pi(N-1)/2$ and for $N$ even $\tilde\alpha = \alpha+\pi N/2$.
If we concentrate on the first two Fourier coefficients of the current we see
from the above strong coupling results of Eq.~(\ref{eq:jlarge}) that:
\begin{eqnarray}
\frac{a_2^o}{a_1^o}
&=&\frac{-12}{J_K L[\tan({\pi}/{2L})+\tan({3\pi}/{2L})]+24}\nonumber\\
\frac{a_2^o}{a_2^e}
&=&\frac{-4}{(1+{1}/{\cos({\pi}/{L^e})})^2}\frac{L^{e}}{L^{o}}.
\label{eq:stronghalf}
\end{eqnarray}
Here the superscripts o and e refer to $N$ odd or even, respectively.
In the limit $L\to\infty$ we see that the strong coupling expansion gives:
\begin{equation}
\frac{a_2^{o}}{a_1^{o}}
\to\frac{-6}{\pi J_K+12}, \ \ \frac{a_2^{o}}{a_2^{e}}\to -1.
\label{eq:thermostronghalf}
\end{equation}
The Fermi liquid theory developed in the preceeding section, 
valid for $1\ll \xi_K\ll L$,
predicts, using Eqs.~(\ref{2}) and (\ref{eq:sin}):
\begin{equation}
\frac{a_2^{o}}{a_1^{o}}
=-\frac{1}{2},\ \
\frac{a_2^{o}}{a_2^{e}}
=-1.\label{eq:flthalf}
\end{equation}
\begin{figure}[t]
\begin{center}
\includegraphics[clip,width=8cm]{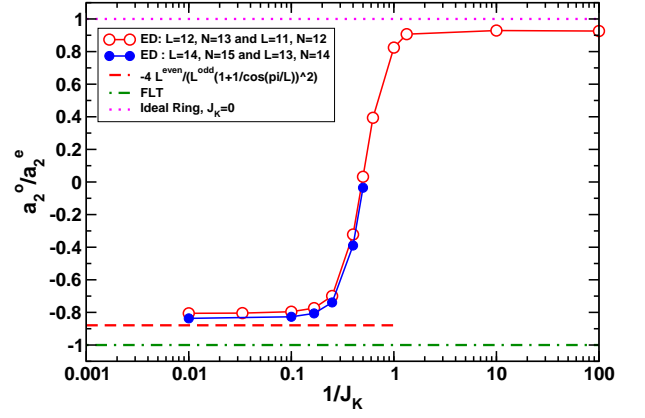}
\caption{The ratio, $a_2^o/a_2^e$, of the second Fourier coefficients of the current for
the SCQD at 1/2-filling, for $N$ odd and $N$ even. $\circ$ indicates ED results with $L=11$, $N=12$  and $L=12$, $N=13$ while
$\bullet$ indicates ED results with $L=13$, $N=14$ and $L=14$, $N=15$. The dashed line is the  strong coupling
perturbative result, Eq.(\ref{eq:stronghalf}), for this ratio with $L^{\rm even}=11, L^{\rm odd}=12$, the dashed dotted line the fermi liquid 
result, Eq.~(\ref{eq:flthalf}), and the dotted line
the ideal ring result, Eq.~(\ref{eq:zerohalf}).
\label{fig:HalfRatOddEven}}
\end{center}
\end{figure}

When $J_K$ is exactly zero we obtain the behavior of an ideal ring, in this case the current has a 
charateristic saw-tooth like form~\cite{ProkofevGogolin,SAprl} where the same Fourier coefficients
can be trivially calculated. One finds at $J_K=0:$
\begin{equation}
\frac{a_2^{o}}{a_1^{o}}
=\frac{1}{2},\ \ 
\frac{a_2^{o}}{a_2^{e}}
=1.
\label{eq:zerohalf}
\end{equation}
Note that this result is dramatically different from the result of Eq.~(\ref{eq:flthalf}),
valid for $1\ll \xi_K\ll L$. The ratio of the Fourier coefficients appear discontinuous in the limit $J_K\to 0$.

We begin by discussing our results for ${a_2^o}/{a_1^o}$ which are shown in
Fig.~\ref{fig:HalfRatOddOdd} for system sizes of $L=12$ and $L=14$, for a range
of different $J_K$. The theoretical strong coupling result of
Eq.~(\ref{eq:stronghalf}) for $L=12$ is shown as the dashed line and differs
only slightly from the $L\to\infty$ result of Eq.~(\ref{eq:thermostronghalf})
shown as the solid line. The numerical results agrees with the strong
coupling results once $J_K>10$. In the opposite limit, $J_K\to 0$, we see
that the numerical results quickly approach the ideal ring result of
Eq.~(\ref{eq:zerohalf}), shown as the dotted line.  This is very reasonable,
since the very limited system sizes of $L=12,14$ certainly no longer
satisfies the equality $L\gg\xi_K$ for Fermi liquid theory to be valid once
$\xi_K$ diverges as $J_K\to 0$. The interesting region is therefore the
intermediate coupling region, $0.1<J_K<10$. A very rapid crossover from the
ideal ring result of 1/2, at small $J_K$, to a negative value is observed, consistent with an
exponentially diverging $\xi_K$. Furthermore, ${a_2^o}/{a_1^o}$ develops a
pronounced plateau (``dip") around $J_K\sim 4$,
reaching negative values, even for these
rather modest system sizes. As we increase the system size from $L=12$ to $L=14$ this
``dip" becomes slightly more pronounced.
This implies that the cross-over between the ideal ring result for this ratio of 1/2, at small $J_K$, to
the $J_K\to\infty$ result of 0, {\it can not} be monotonic even in the thermodynamic limit.
We take this to be indicative of the validity of
the Fermi Liquid theory. We expect that numerical results for
${a_2^o}/{a_1^o}$ in the thermodynamic limit would roughly follow the strong
coupling result given by Eq.~(\ref{eq:thermostronghalf}) (the solid line in
Fig.~\ref{fig:HalfRatOddOdd}) out to $J_K\sim 0.1$ and then attain the value -1/2
rather quickly, jumping discontinuously for $J_K=0$, in accordance with the
Fermi Liquid theory.  
In the subsequent section, where we discuss our results
away from half-filling, we therefore exclusively focus on the value of
$J_K=4$ where the ``dip" occurs, since this would appear to be the most promising value for $J_K$ for
observing FLT behavior with the available system sizes.

Results for $a_2^o/a_2^e$ are shown in Fig.~\ref{fig:HalfRatOddEven}. In this
case the behavior as a function of $J_K$ is monotonic and the evidence for a
regime described by FLT perhaps less obvious. The dotted, dashed and
dashed-dotted lines are the ideal ring results of Eq.~({\ref{eq:zerohalf}), the
strong coupling result of Eq.~(\ref{eq:stronghalf}) and the FLT result of
Eq.~(\ref{eq:flthalf}). Again an extremely rapid crossover is seen at
intermediate values of $J_K$. Furthermore, as the system size is increased
from $L=12$ to $L=14$ this crossover moves to smaller values of $J_K$
approaching the predictions of Fermi Liquid theory.

\subsection{Off Half-filling\label{sec:offhalf}}
\begin{figure}[t]
\begin{center}
\includegraphics[clip,width=8cm]{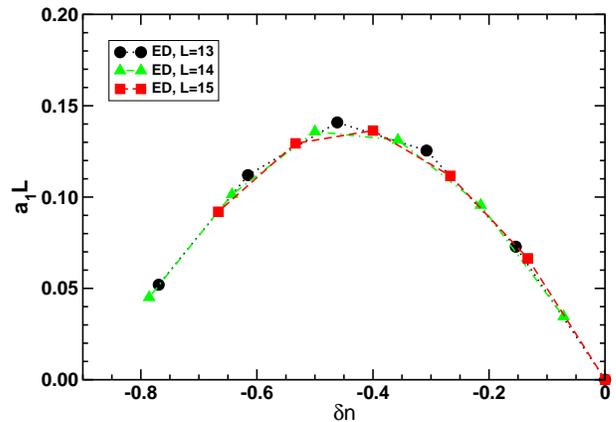}
\caption{The first Fourier coefficient scaled by $L$, $a_1^e L$, of the current for
the SCQD away from 1/2-filling, for $N$ even as a function of $\delta n=(N-1)/L-1$. In all cases $J_K=4.0$. $\circ$ indicates
ED results with $L=13$ and $N=14,12,10,8,6,4$, $\triangle$ indicates ED results with $L=14$ and $N=14,12,10,8,6,4$, finally
$\bullet$ indicates ED results with $L=15$ and $N=16,14,12,10,8,6$. The lines are guides to the
eye.
\label{fig:a1Nevenoffhalf}}
\end{center}
\end{figure}
Next, we turn to a discussion of our results for off half-filling.
With our definitions of $N$ and $L$ we define, as elsewhere, $\delta n = (N-1)/L -1$.
A typical Hilbert space size is  $64,414,350$ for $(L=15,N=14)$  after symmetry reductions.
In Ref~\onlinecite{prl.persistent} strong coupling perturbative results for
the current past the SCQD away from half-filling were also given. At large
$J_K$ it was found that for $N$ odd and $N$ even:
\begin{eqnarray}
\lefteqn{j_oL/e\approx\frac{32}{9J_K^2}\sin (\tilde \alpha )}&&\nonumber\\
&&\times\left[\sin \frac{\pi (N-1)}{2L}\tan \frac{\pi}{2L}
-\sin \frac{3\pi (N-1)}{2L}\tan \frac{3\pi}{2L}\right]\nonumber\\
\lefteqn{j_eL/e \approx  \frac{32}{9J_K^2} \sin (\tilde \alpha )}&&\nonumber\\
&&\times\left[ \frac{\cos (\pi (N-1)/(2L))}{\cos (\pi /(2L))}-
\frac{\cos (3\pi (N-1)/(2L))}{\cos (3\pi/(2L))}\right]\nonumber\\
\label{eq:scqdoff}
\end{eqnarray}
This results predicts that as $L\to\infty$, $\delta n \to 0$, $j_oL/e$ is changed
from the value at half-filling, Eq.~(\ref{eq:jlarge}), only by a term proportional
to $-(\delta n)^2\sin(\tilde\alpha)/(J_K^2L)$. If we only work to linear order in $\delta n$
this correction can be neglected.
However in this limit, $j_e L/e\sim \sin(\tilde\alpha)64\pi\delta n/9 J_K^2$, that is, a $\sin(\tilde\alpha)$ term  proportional
to $\delta n$. At half-filling this term is clearly 0 for $j_e$, and the $\sin(2\tilde\alpha)$ term
in Eq.~(\ref{eq:jlarge}) dominates. 
We note that this limit of the strong coupling results agrees with the previously
developed Fermi liquid theory that, in the limit $\delta n\to 0$, predicted no change in $j_o$
but the generation of a $\sin(\tilde\alpha)$ term proportional to $\delta n/L$ for $j_e$,
Eq.~(\ref{lambdan}), (\ref{jPH}). In the following, we therefore focus exclusively on the first Fourier
coefficient, $a_1$, determining its dependence on $\delta n$.
\begin{figure}[t]
\begin{center}
\includegraphics[clip,width=8cm]{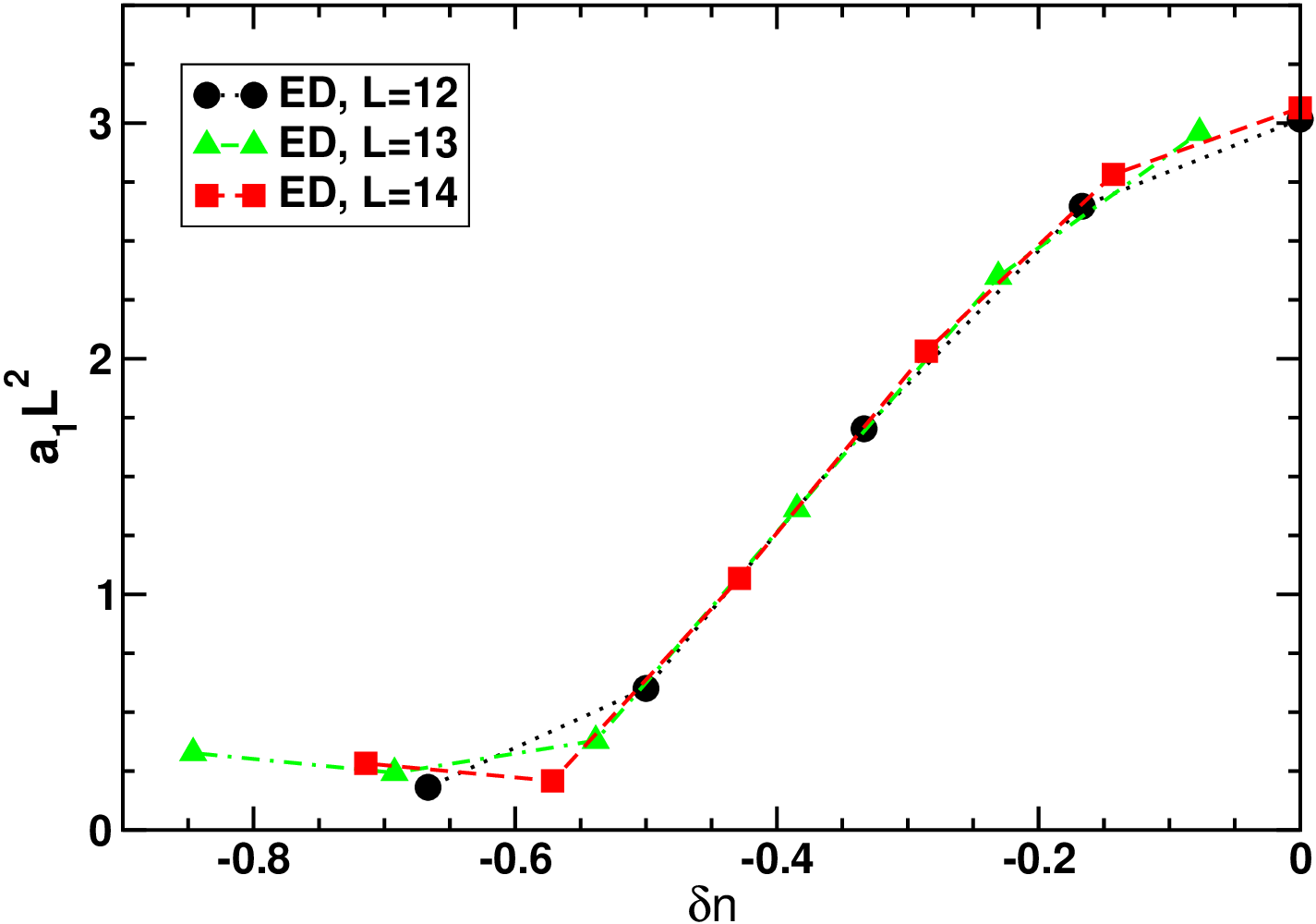}
\caption{The first Fourier coefficient scaled by $L^2$, $a_1^o L^2$, of the current for
the SCQD away from 1/2-filling, for $N$ odd as a function of $\delta n=(N-1)/L-1$. In all cases $J_K=4.0$. $\circ$ indicates
ED results with $L=12$ and $N=13,11,9,7,5,3$,
$\triangle$ indicates ED results with $L=13$ and $N=13,11,9,7,5,3$, finally
$\bullet$ indicates ED results with $L=14$ and $N=15,13,11,9,7,5$. The lines are guides to the
eye. 
\label{fig:a1Noddoffhalf}}
\end{center}
\end{figure}

Our results for $a_1^e L$ for $N$ even are shown in Fig.~\ref{fig:a1Nevenoffhalf} for system sizes $L=13,14,15$
for a range of $\delta n$. All results are for an intermediate coupling of $J_K=4$, the most promising
coupling to show clear indications of FLT behavior for our limited system sizes. Two conclusions are
immediately evident; since the results fall on a single curve this term in the current is indeed proportional
to $1/L$ as predicted by theory. Secondly, for small $\delta n$, $a_1L$ increases approximately  linearly with $\delta n$,
again consistent with the theoretical predictions. At larger $\delta n$ the results strongly deviate from linear
behavior as one would expect. 

Finally, we show results for $a_1^o L^2$ for $N$ odd in Fig.~\ref{fig:a1Noddoffhalf} for system sizes of $L=12,13,14$,
for a range of $\delta n$. The results are again for an intermediate coupling of $J_K=4$. As was the case for $a_1^e L$,
the results for different system sizes again follow a single curve, validating the scaling $a_1^o\sim 1/L^2$ predicted
by theory. A term proportional to $1/L$ can definitely be excluded for this Fourier coefficient.
We also see that, for $\delta n\to 0$, our results are consistent with deviations from the result at half-filling
being of higher than linear order in $\delta n$, although a definite conclusion is hard to obtain due to the limited
system sizes available.

\section{Finite Temperature Conductance\label{sec:conductance}}
We now 
calculate the conductance for a quantum dot, side-coupled 
to infinitely long leads, 
at a low finite temperature, $T<<T_K$.  
We do this using our Fermi liquid Hamiltonian, of Eq. (\ref{flt}). We choose $L$ and $L/2$ 
even, for convenience but we are now taking the $L\to \infty$ limit 
so this choice is immaterial.  Working with left-movers only, and taking into account 
that the point $x=L$ corresponds to $x=0^-$, $x=0$ to $0^+$, we write the Fermi liquid 
interaction as:
\bea H_{int}={-8\pi \over 3T_K}:[\psi^\dagger 
{\vec \sigma \over 2} \psi]^2: 
-{4i\over 3 T_K}:\psi^\dagger 
\overleftrightarrow{d\over dx}\psi :
+\hbox{const}.\label{FLHC}\eea
Here 
\begin{equation} \psi \equiv {1\over \sqrt{2}}\left[\psi_L(0^+)-\psi_L(0^-)e^{-i\alpha}
\right],\label{psidef}\end{equation}
the $:\ :$ denotes normal ordering and we define:
\begin{equation} f(x)\overleftrightarrow{d\over dx}g(x)\equiv f{dg\over dx}-
{df\over dx}g.\end{equation}
The second, derivative term in Eq. (\ref{FLHC}), arises from 
a point splitting procedure. The second term is referred to as the elastic 
part of the interaction, corresponding to a single electron impuirity scattering 
process, while the first term is referred to as the inelastic part, 
corresonding to an electron-electron interaction at the origin. 
(For a derivation of the Fermi liquid interaction 
in this form, see [\onlinecite{AL}], but note that the fermion fields are 
defined with an unconventional normalization there so that they are larger by 
a factor of $\sqrt{2\pi}$.) 
Now the phase $\alpha$, is regarded as a time-dependent vector potential, related to the 
potential energy drop across the quantum dot by:
\begin{equation} d\alpha /dt= -\int_{0^-}^{0^+} dx E(x)=\Delta V.\label{V}\end{equation}
(In this section we set the electron charge, $e=1$, reinstating it 
at the end.) 
The conductance through a quantum dot, treating the leads as ballistic, is insensitive 
to where the electric field is applied so we have chosen, for convenience, to
apply it right at the junction, over a region of width of order $\xi_K$.

We write the current operator as the rate of change of one-half the 
difference of the total number of electrons on 
the right hand side ($x>0$) of the system minus the total number on the left hand side:
\begin{equation} I= (1/2)d\hat (N_+-\hat N_-)/dt = (1/2)i[H,\hat N_+-\hat N_-],\end{equation}
where 
\begin{equation} \hat N_+\equiv \int_0^\infty [\psi^\dagger_L(x)\psi_L(x)+\psi^\dagger_R(x)\psi_R(x)],\end{equation}
and similarly for $\hat N_-$. 
As discussed above, we can use the perfectly reflecting strong-coupling b.c. of Eqs. (\ref{bc0}), 
(\ref{bcL}))
to rewrite the left and right movers at $x>0$ in terms of left-movers only on the entire real line. 
Thus we define:
\bea \psi_{L+}(x) &\equiv& \psi_L(x),\ \  (x>0)\nonumber \\
&=& \psi_R(-x) \ \  (x<0).\eea
Here the subscript $+$ indicates that this field derives from the $x>0$ region.  Similarly, 
we define another left-moving field on the entire real line, $\psi_{L-}(x)$ which 
derives from the region $x<0$:
\bea \psi_{L-}(x) &\equiv& \psi_L(x),\ \  (x<0)\nonumber \\
&=& \psi_R(-x) \ \  (x>0).\eea
In this notation:
\begin{equation} \hat N_\pm = \int_{-\infty}^\infty \psi^\dagger_{L\pm}(x)\psi_{L\pm}(x)\end{equation}
and the operator appearing in the Fermi liquid interaction, 
defined in Eq. (\ref{psidef}) becomes:
\begin{equation} \psi \equiv{1\over \sqrt{2}}\left[\psi_{L+}(0)-\psi_{L-}(0)e^{-i\alpha}
\right].\end{equation}

We wish to calculate $<I>$ in linear response to a time dependent 
vector potential $\alpha (t)$. At this point, there are two 
ways of proceeding. We may use the expression $I=(1/2)d(\Delta \hat N)/dt$ 
where
\begin{equation} \Delta \hat N\equiv \hat N_+-\hat N_-,\end{equation}
and ultimately relate the conductance to the Green's function 
of $\Delta \hat N$, or we may calculate explicitly 
the commutator $[H,\Delta \hat N] $ and relate the conductance to 
the Green's function of that operator. In both 
approaches the calculation must be carried out to second order 
in the Fermi liquid interaction and the amount of work is 
roughly the same either way. However, it is actually much 
more convenient to work with $d \Delta \hat N/dt$ because the 
needed Green's function can be related to the ${\cal T}$-matrix, 
which was calculated previously.\cite{Nozieres,AL} This 
approach was used in [\onlinecite{GP}] for instance, to 
calculate the conductance through an {\it embedded} quantum dot. 
The calculation is very similar in the side-coupled case and 
we may obtain the answer by only a slight modification of 
their result. This approach is used in the rest of 
this section.  It is also of some interest to calculate 
the conductance using the commutator method and this is 
done in  Appendix A, yielding precisely the same result. 

In the $d\Delta \hat N/dt$ method, we use the fact that, to $O(\alpha )$, 
$H=H(0)+\alpha (t)I$, to obtain the conductance:
\begin{equation} C=\lim_{\omega \to 0} {1\over \omega}\int_0^\infty dt e^{i\omega t}<[I(t),I(0)]>.\end{equation}
Integrating by parts:
\begin{equation} C=(1/4)\lim_{\omega \to 0}\omega \int_0^\infty d\omega e^{i\omega t}
<[\Delta \hat N(t),\Delta \hat N(0)]>,\end{equation}
where the retarded Green's function is evaluated at $\alpha =0$. It is now 
convenient to go to the even-odd basis:
\begin{equation} \psi_{e/o}={1\over \sqrt{2}}(\psi_{L+}\mp \psi_{L-}).\end{equation}
Since the interactions only involve $\psi_e$, $C$ factorizes into 
a free Green's function for $\psi_o$ multiplied by the non-trivial Green's function 
for $\psi_e$. This latter Green's function can be expressed in terms of 
the ${\cal T}$-matrix, giving a formula for the conductance:\cite{PG}
\begin{equation} C={e^2\over h}\sum_s \int d\epsilon (-df/d\epsilon )
[-\pi \nu \hbox{Im} {\cal T}_s (\epsilon )],\label{C-T}\end{equation}
where $s=\pm 1$ labels the fermion spin, 
$f(\epsilon )$ is the Fermi distribution function at temperature $T$ and 
$\nu$ is the density of states. As shown in [\onlinecite{GP}] following 
[\onlinecite{Nozieres}, \onlinecite{AL}], for the embedded quantum dot:
\begin{equation} -\pi \nu T_s (\epsilon )={1\over 2i}\left[\exp [2i\delta_s (\epsilon )]-1\right]
+\exp [2i\delta_s (\epsilon )][-\pi \nu \tilde{\cal T}_{in}(\epsilon )],\end{equation}
where 
\begin{eqnarray} \delta_s (\epsilon )&=& s\pi /2+\tilde \delta (\epsilon)\nonumber \\
\tilde \delta (\epsilon ) &=& \omega /T_K \nonumber \\
-\pi \nu \tilde T_{in}(\epsilon )&=&i\frac{(\epsilon^2+\pi^2T^2)}{2T_K^2}.\end{eqnarray}
$\tilde \delta (\epsilon )$ and $\tilde T_{in}(\epsilon )$ are the 
phase shift and inelastic part of the ${\cal T}$-matrix calculated in 
perturbation theory in the Fermi liquid interactions.  The extra 
$s\pi /2$ term in $\delta_s$ arises from the $\pm \pi /2$ phase shift 
characterizing the strong coupling fixed point. This $\pm \pi /2$ phase 
shift reflects the perfect transmission ($C=2e^2/h$) at $T=0$ 
for the embedded quantum dot.  On the other hand, for 
the side-coupled quantum dot, $C=0$ at the zero temperature fixed point, 
implying the absence of this extra phase shift.  Thus, for the side-coupled 
quantum dot we have:
\begin{equation} -\pi \nu T_s (\epsilon )={1\over 2i}\left[\exp [2i\tilde \delta_s (\epsilon )]-1\right]
+\exp [2i\tilde \delta_s (\epsilon )][-\pi \nu {\cal T}_{in}(\epsilon )].\end{equation}
To order $\omega^2$, $T^2$, the difference between ${\cal T}$-matrices for 
the embedded and side-coupled quantum dots is just a change in sign in 
the first and third terms.  This implies:
\bea -\pi \nu \hbox{Im}{\cal T} (\epsilon )&=&1-\frac{(3\epsilon^2+\pi^2T^2)}{2T_K^2}\ \  (\hbox{embedded})\nonumber \\
-\pi \nu \hbox{Im}{\cal T} (\epsilon )&=&\frac{(3\epsilon^2+\pi^2T^2)}{2T_K^2}\ \  (\hbox{side-coupled})\eea
Inserting these expressions into Eq. (\ref{C-T}) for the conductance, and restoring 
$\hbar$ and $e$ which were previously set to one, gives:
\bea C&=&{2e^2\over h}\left[ 1-{\pi^2T^2\over T_K^2}\right] ,\ \  
\ \ (\hbox{embedded})\nonumber \\
&=&{2e^2\over h}{\pi^2T^2\over T_K^2}\ \ (\hbox{side-coupled})\label{condfin}\eea
 
 We may now write universal Wilson-type ratios between the low temperature conductance, $C$,  
through a side-coupled quantum dot and the zero temperature persistent current, $j_e$, $j_o$, through 
the same quantum dot inserted into a ring of size $L\gg \xi_K$:
\bea 
{L^2j_e/ev_F^2\over \sqrt{hC/(2e^2T^2)}}&
\to &{1\over 4}\sin 2\tilde \alpha \nonumber \\
{L^2j_o/ev_F^2\over \sqrt{hC/(2e^2T^2)}}&
\to&{1\over 2}\sin\tilde \alpha -{1\over 4}\sin 2\tilde \alpha .
\eea

When PH symmetry is broken the conductance also gets a contribution of 
second order in $\lambda$ the coupling constant in the term, $H_2$ 
of the effective Hamiltonian, given in Eq. (\ref{H2}). The simplest 
way of evaluating this contribution is to observe that, since 
the Hamiltonian is non-interacting, ignoring the other 
term $H_{int}$ in Eq. (\ref{flt}), we can evaluate the conductance 
at zero temperature 
using the Landauer formula, $C=(2e^2/h)T_r$ where $T_r$ is the transmission 
probability through the quantum dot. For small $\lambda$, $T_r=\lambda^2$. 

Including both terms, the conductance takes the form at low temperatures:
\begin{equation} C = \left[ {\pi^2T^2\over T_K^2}+\lambda^2\right] {2e^2\over h}.
\end{equation}
At sufficiently low $T$ the second, non-universal term, 
arising from PH symmetry breaking always dominates. However 
provided that the PH symmetry breaking (and, in particular, 
the direct tunnelling across the quantum dot) is small, 
the first, universal, term dominates for:
\begin{equation} \lambda T_K<<T<<T_K.\end{equation}
Note that this situation is very analogous to that 
for the persistent current, with $T$ replaced by $v_F/L$.

\section{Conclusion}
We have developed a Fermi liquid theory of the persistent current in a side coupled quantum dot.
This theory should correctly describe the limit $a\ll \xi_K\ll L$ where $a$ is a microscopic 
length scale. Numerical results are
largely in agreement with the existence of a regime correctly described by this Fermi liquid picture.
However, probably due to the limited system sizes available and the requirement that $L\gg\xi_K$, 
the numerical evidence cannot be described as strong. Our Fermi liquid theory confirms
the existence of a term in the persistent current proportional to $\sin(\tilde\alpha)\delta n/J_K^2L$ for $N$ even, 
absent at half-filling.  We have also calculated the conductance through a side-coupled quantum 
dot at low $T\ll T_K$ and calculated a universal ``Wilson'' ratio relating the conductance to the 
persistent current.

\appendix
\section{Conductance by Commutator Method}
In this appendix we repeat the calculation of the conductance, writing the 
current as $I=i[H,\Delta \hat N]/2$.  This provides a check on the previous 
Fermi liquid calculations involving the ${\cal T}$-matrix in [\onlinecite{Nozieres,AL,GP,PG}] 
and also provides an instructive example of how universal information can be 
extracted from a cut-off dependent result. 

Since $[\hat N_++\hat N_-,H]=0$, we may equivalently use $I=i[H,\hat N_+]$.
Note that $\hat N_+$ commutes with the non-interacting part of the Hamiltonian which 
is simply that of free fermions with perfectly reflecting b.c.'s at the origin, since 
this Hamiltonian does not transmit any electrons between left and right sides of the system. 
The commutator of $\hat N_+$ with $H_{int}$ is readily calculated.  Since all fields in the 
commutator are left-movers sitting at $x=0$, we drop the $L$ subscript and the $(0)$ argument 
for simplicity. The result is:
\begin{widetext}
\bea I=i[H_{int},N_+] &=& -{2\pi i\over 3T_K}\{[e^{-i\alpha}\psi_+^\dagger{\vec \sigma \over 2}\psi_- -h.c.],
[\psi_+-e^{-i\alpha}\psi_-]^\dagger{\vec \sigma \over 2}[\psi_+-e^{-i\alpha}\psi_-]\}\nonumber \\
&& +{1\over 2T_K}\left[\psi_+^\dagger{d\over dx}\psi_-e^{-i\alpha}-{d\over dx}\psi^\dagger_+\psi_-e^{-i\alpha}
+h.c.\right]
.\eea
\end{widetext}
We note that this is, equivalently, $I=dH_{int}/d\alpha$. 
We expand the current operator up to first order in the vector potential, $\alpha$:
\begin{equation} I = I_0+\alpha I_1 + O(\alpha^2).\end{equation}
Here:
\bea I_0 &=& {4\pi i\over 3T_K}[(\psi_+^\dagger {\vec \sigma \over 2}\psi_-)^2-\psi_+^\dagger {\vec \sigma \over 2}\psi_-\cdot 
(\vec J_++\vec J_-)-h.c.]\nonumber \\&&
+{1\over 2T_K}\left[\psi_+^\dagger\overleftrightarrow{d\over dx}\psi_- +h.c. \right]
\equiv I_{in}+I_{el},
\eea
where
\begin{equation} \vec J_{\pm} \equiv \psi_{\pm}^\dagger {\vec \sigma \over 2}\psi_{\pm}\end{equation}
and we have defined the inelastic and elastic terms in the current operator 
which are, respectively, quartic and quadratic in fermion operators. 
The precise form of $I_1$ will not be needed. 
The conductivity is obtained from $<I>$, calculated to first order in $\alpha$. 
Thus we obtain:
\begin{equation} I(\omega ) = [G_R(\omega)+<I_1>]\alpha (\omega ),\end{equation}
where $G_R$ is the retarded Green's function:
\begin{equation} G_R(\omega ) = -i\int_0^\infty e^{i\omega t}<[I_0(t),I_0(0)]>_T.\end{equation}
It can be seen that $I=0$ when $\alpha (t)$ is a constant, independent of $t$. 
A non-zero $I$ in this case would correspond to a persistent current, but this 
must vanish for infinite $L$ since a constant phase $\alpha$ in $H_{int}$ can 
be eliminated by a gauge transformation in that limit. Thus we conclude that
$<I_1>=-G_R(0)$, so that:
\begin{equation} I(\omega )=[G_R(\omega )-G_R(0)]\alpha (\omega ).\end{equation}
From Eq. (\ref{V}), we see that:
\begin{equation} I(\omega )=C(\omega )\Delta V(\omega ),\end{equation}
where the conductance, $C(\omega )$, is given by:
\begin{equation} C(\omega ) = {i\over \omega }[G_R(\omega )-G_R(0)],\end{equation}
where $\omega_n=2\pi nT$ and $\beta \equiv 1/T$.  (We set Boltzmann's constant 
equal to one.) 
The dc conductance is:
\begin{equation} C=\lim_{\omega \to 0}{i\over \omega }[G_R(\omega )-G_R(0)].\label{CDC}\end{equation}
We calculate the retarded Green's function of $I_0$ by analytic continuation 
from the imaginary time, Matsubara Green's function:
\begin{equation} \GG(i\omega_n )=-\int_0^\beta d\tau e^{i\omega_n\tau}<I_0(\tau )I_0(0)>.\end{equation}
The free fermion Matsubara 
Green's function, at late times, using the same normalization as in Eq. (\ref{prop}), that 
was used in the calculation of the persistent current is:
\begin{equation} <\psi^{\dagger \alpha}_i(\tau )\psi_{\gamma j}(0)>\to {\delta^\alpha_\gamma 
\delta_{ij}\over 2\beta \sin (\pi \tau /\beta )},\end{equation}
where $i$, $j=\pm$ labels right and left sides of the junction, $x=0^{\pm}$ 
and $\beta$ is the inverse temperature. Using Wick's theorem, the identity:
\begin{equation} \sum_{a,b}[tr \sigma^a\sigma^b~tr\sigma^a\sigma^b-tr\sigma^a\sigma^b\sigma^a\sigma^b]=18,\end{equation}
and collecting the various terms, we obtain the long-time behavior of the Matsubara Green's function:
\begin{equation} <I_{in}(\tau )I_{in}(0)>\to {16\pi^2\over T_K^2}{1\over (2\beta )^4 \sin^4 (\pi \tau /\beta )}.\end{equation}

To evaluate $<I_{el}(\tau )I_{el}(0)>$, we use the derivatives of the free fermion Green's function:
\begin{eqnarray}
{d\over dx}\left\{ {1\over \sin [\pi (\tau + ix)/\beta ]}\right\} &=&
 -{i\pi \cos [\pi (\tau +ix)/\beta ]\over \beta \sin^2 [\pi (\tau +ix)/\beta ]},
\nonumber \\
{d^2\over dx^2} {1\over \sin [\pi (\tau + ix)/\beta ]} &=&
-{\pi^2\over \beta^2}{1+\cos^2[\pi (\tau + ix)/\beta ]\over \sin^3 [\pi (\tau + ix)/\beta ]}.\nonumber\\
\end{eqnarray}
The terms proportional to $\cos^2(\pi \tau /\beta )$  cancel leaving:
\begin{eqnarray} 
<I_{el}(\tau )I_{el}(0)>&\to& {8\pi^2\over T_K^2}{1\over (2\beta )^4\sin^4(\pi \tau /\beta)}\nonumber\\
&=&(1/2) <I_{in}(\tau )I_{in}(0)>.
\end{eqnarray}
We now consider the Fourier transform of $G(\tau )$ at low frequencies and temperatures.  The 
analytic continuation of $\GG(i\omega_n)$ ($\omega_n\equiv 2\pi n/\beta$) to real frequencies gives the retarded Green's function: 
\begin{equation} \GG(i\omega_n\to \omega +i\delta)=G_R(\omega ),\end{equation}
where $\delta$ is a postive infinitesimal. 
We will find that 
\begin{equation} \GG(i\omega_n)\to A+B|\omega_n| + O(\omega_n^2),\end{equation}
for $\omega_n<<D$. The analytic continuation gives:
\begin{eqnarray} |\omega_n|&=&\omega_n\int_{-\infty}^\infty {d\tau \over 2\pi i}{e^{i\omega_n\tau}
-e^{-i\omega_n\tau}\over \tau -i\epsilon}\nonumber\\
&\to& (-i\omega)
\int_{-\infty}^\infty {d\tau \over 2\pi i}{e^{(i\delta -\omega )\tau}
-e^{-i(\delta -i\omega )\tau}\over \tau -i\epsilon}\nonumber\\
  &=& -i\omega 
.\label{omegacon}\end{eqnarray}
$G(\omega_n)$, which is an even function of $\omega_n$, in our assumed particle-hole 
symmetric model, contains non-universal ultraviolet cut-off dependent terms. Introducing 
a cut-off, $\tau_0$ with dimensions of time, and of order $1/D$, we find, at low $\omega_n$:
\begin{eqnarray} 
\lefteqn{\GG(i\omega_n)\to}&&\nonumber\\
&&{1\over T_K^2}\left[{a\over \tau_0^3}+b{1\over \tau_0\beta^2}
+c{|\omega_n|\over \beta^2}+d{\omega_n^2\over \tau_0}+e|\omega_n|^3+\ldots \right],
  \nonumber\\
  \label{loww}
\end{eqnarray} 
where $a,b,c,d,e$ are dimensionless numbers. 
While the terms of O$(\omega_n^0)$ and $(\omega_n^2)$ are non-universal and cut-off 
dependent, we expect that the terms $\propto |\omega_n|$ and $|\omega_n|^3$ are not. This is related to the 
fact that only these terms are  singular functions of $\omega_n$.  This singularity 
arises from the universal long-time behavior of $G(\tau )$. Note that the 
dc conductance is completely determined by the universal $c$-term $\propto |\omega_n|$ 
and so is a universal quantity in the sense that it is independent of the 
details of the ultraviolet cut-off.
To verify these assertions, we consider a particularly simple ultraviolet cut-off.  
The $\tau$ integral defining the Fourier transform of 
the current Green's function is restricted to $\tau_0<\tau <\beta - \tau_0$
Thus we begin with:
\begin{equation} \GG(i\omega_n)\equiv {-24\pi^2\over T_K^2(2\beta )^4}\int_{\tau_0}^{\beta -\tau_0}
{d\tau e^{i\omega_n\tau}\over \sin^4(\pi \tau /\beta )}.
\label{int-cutoff}\end{equation} 
This integral can be straightforwardly 
developed in an expansion in $\tau_0/\beta$, which has the form of Eq. (\ref{loww}).
This can be conveniently done by deforming the $\tau$-integral into the complex plane. 
The original integral is equal to the sum of an integral along the straight line from $\tau = i\tau_0$ 
to $\tau = i\tau_0+\beta$ plus the integral on two quarter-circle contours from  
$\tau = \tau_0$ to $\tau = i\tau_0$ and from $\beta + i\tau_0$ to $\tau = \beta -\tau_0$, which we label $C_1$ and $C_2$ respectively.  [See fig. 
(\ref{fig:contour}).]
\begin{figure}
\begin{center}
\includegraphics[clip,width=8cm]{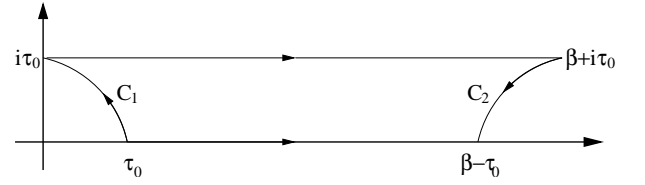}
\caption{Contour used to evaluate the integral in Eq. (\ref{int-cutoff}).}
\label{fig:contour}
\end{center}
\end{figure}
The integral along the straight line can be written:
\begin{equation} K=\int_0^\beta {d\tau e^{i\omega_n(\tau +i\tau_0)}\over \sin^4[\pi (\tau +i\tau_0)/\beta ]}.\label{int_st}\end{equation}
We may now Taylor expand the denominator:
\begin{equation} {1\over \sin^4[\pi (\tau +i\tau_0)/\beta ]}=16e^{4\pi i(\tau +i\tau_0)/\beta}
\sum_{m=0}^\infty a_me^{2\pi im(\tau +i\tau_0)/\beta}.\end{equation}
Note that the sum converges due to the factors $e^{-2\pi \tau_0/\beta}$. Here the $a_m$
are the binomial coefficients:
\begin{equation}  a_m= (m+1)(m+2)(m+3)/6.\end{equation}
We now integrate over $\beta$ term by term. The result is:
\begin{equation} \int_0^\beta d\tau e^{i[\omega_n+(2+m)(2\pi /\beta )]\tau}=\beta \delta_{n,-(2+m)}.\end{equation}
(Recall that $\omega_n\equiv 2\pi n/\beta$.)  Thus we obtain:
\begin{equation} K=-{4\beta^2|\omega_n|\over 3\pi}\left[
1-\left({\beta \omega_n \over 2\pi}\right)^2\right]
\theta (-\omega_n).\end{equation}
Alternatively, the integral of Eq. (\ref{int_st}) can be evaluated 
by introducing the complex variable $z\equiv e^{i2\pi (\tau+i\tau_0)/\beta}$. 
The integration contour for $z$ is a circle of radius $e^{-2\pi \tau_0/\beta}<1$.
This contour encloses a pole of order $|n|-1$ when $n<-1$. 

We also 
must perform the integration over the two quarter-circle contours:
\begin{equation} K_{12}\equiv \sum_{i=1}^2
\int_{C_i}dz{e^{i\omega_nz}\over \sin^4(\pi z/\beta)}.\end{equation}
On $C_1$ we write $z\equiv \tau_0e^{i\theta}$ where $\theta$ goes from 
$0 \to \pi /2$ along the contour. On $C_2$ we write $z=\beta +\tau_0
e^{i\theta}$ where now $\theta$ goes from $\pi /2\to \pi$. Since 
the integrand is invariant under a translation: $z\to z+\beta$, 
we may combine these two terms to write:
\begin{equation} K_{12}=i\tau_0\int_0^\pi d\theta {e^{i\theta}\exp 
[i\omega_n\tau_0e^{i\theta}]\over \sin^4(\pi \tau_0e^{i\theta}/\beta )}.\end{equation}
Next we expand $\exp [i\omega_n\tau_0e^{i\theta}]$ in powers of 
$\omega_n\tau_0$ and expand $\sin^{-4}(\pi \tau_0e^{i\theta}/\beta )$
in powers of $\tau_0/\beta$, giving:\begin{widetext}
\begin{equation} K_{12}=i\tau_0\left({\beta \over \pi \tau_0}\right)^4\int_0^\pi 
e^{i\theta}e^{-4i\theta}\left[1+i\omega_n\tau_0e^{i\theta}-
{\omega_n^2\tau^2\over 2}e^{2i\theta}-{i\omega_n^3\tau_0^3\over 6}e^{3i\theta}
+\ldots \right]\left[1-{\pi^2 \tau_0^2\over 6\beta^2}e^{2i\theta}
+{\pi^4\tau_0^4\over 120 \beta^4}e^{4i\theta}+\ldots \right]^{-4}.\end{equation}
\end{widetext}
We Taylor expand, collect terms, and integrate term by term using:
\bea \int_0^\pi d\theta e^{im\theta}&=&\pi \ \  (m=0)\nonumber \\
&=& 0\ \  (m \ \hbox{even},\  m\geq 2)\nonumber \\
&=& 2i/m,\ \  (m \ \hbox{odd}).\eea
Note that non-zero terms from the expansion must be proportional to 
 $e^{3i\theta}$ or else $e^{2im\theta}$ to give 
a non-zero contribution. The first type of terms are odd functions of 
$\omega_n$, proportional to $\omega_n$ and $\omega_n^3$.  All other 
terms are even functions of $\omega_n$. Keeping only the terms 
which are non-vanishing as $\tau_0\to 0$ gives:
\begin{equation} K_{12}\approx {-2\omega_n\beta^2\over 3\pi}+{\omega_n^3\beta^4\over 6\pi^3}
+{2\beta^4\over 3\pi^4\tau_0^3}-{\omega_n^2\beta^4\over \pi^4\tau_0}
+{4\beta^2\over 3\pi^2\tau_0}+\ldots \end{equation}
The omitted terms are all even functions of $\omega_n$ and vanish 
as $\tau_0\to 0$. 

Note that when we add $K+K_{12}$ we regain an even function of $\omega_n$, 
as we must:
\begin{eqnarray} K+K_{12}&=&{-2\beta^2\over 3\pi}|\omega_n|\left[1-
\left({\beta \omega_n\over 2\pi}\right)^2\right]\nonumber\\
&+&{2\beta^4\over 3\pi^4\tau_0^3}-{\omega_n^2\beta^4\over \pi^4\tau_0}
+{4\beta^2\over 3\pi^2\tau_0}+\ldots .\end{eqnarray}
The first two terms are singular functions of $\omega_n$ 
at $\omega_n=0$.  All remaining terms are non-singular, even 
powers of $\omega_n$. The singular terms are cut-off independent, 
unlike the non-singular ones. 

Thus we may write the imaginary frequency Green's function for the current 
operator as:
\begin{equation} \GG(i\omega_n)\to \GG(0)+{\pi T^2|\omega_n|\over T_K^2}+O(\omega_n^2).\end{equation}
The analytic continuation to real frequency is straightforward, using 
Eq. (\ref{omegacon}):
\begin{equation} G_R(\omega ) \to G_R(0)-{i\pi T^2\omega \over T_K^2}+O(\omega^2).\end{equation}
Thus, we obtain the dc conductance from Eq. (\ref{CDC}):
\begin{equation} C = {\pi T^2\over  T_K^2}\to {2e^2\over h}{\pi^2T^2\over T_K^2},\end{equation}
the same result obtained from the derivative method, Eq. (\ref{condfin}).
 (A factor of $e^2/\hbar$, previously 
set equal to one was inserted in the last step.) 

In Sec. V, we evaluated the conductance from the P-H symmetry violating 
tunneling term using the Landauer formalism. 
It is instructive to evaluate the conductance instead using the Kubo formula, and the commutator approach, 
as done earlier in this Appendix. In this case, the new term in the ($\alpha =0$) current operator is:
\begin{equation} \delta I_0 = i\lambda e^{i\pi (N-1)/2}\psi^\dagger_+\psi_- + h.c.\end{equation} 
The Green's function for $I_0$, for large imaginary times, now picks up a correction:
\begin{equation} \delta <I_0(\tau )I_0(0)>=4\lambda^2{1\over (2\beta )^2\sin^2(\pi \tau /\beta )}.\end{equation}
We may evaluate the Fourier transform using the same methods as above.  However, 
we focus immediately on the $T=0$ limit, for simplicity:
\begin{equation} \delta \GG (\omega_n)=-4\lambda^2\int_{-\infty}^\infty 
{d\tau e^{i\omega_n\tau}\over (2\pi \tau )^2}.\label{deltaG}\end{equation}

This integral requires an ultraviolet cut off, as before. Cutting off the 
integral at $|\tau |>\tau_0$, gives:
\begin{equation} \delta \GG = {-2\lambda^2\over \pi^2}\int_{\tau_0}^\infty {d\tau \over \tau^2}
\cos (\omega_n\tau ) .\end{equation}
It is convenient to integrate by parts, giving:
\begin{equation} \delta \GG = {-2\lambda^2\over \pi^2\tau_0}+{2\lambda^2\omega_n\over \pi^2}
\int_{\tau_0}^\infty {d\tau \over \tau}
\sin (\omega_n\tau ).\end{equation}
[We used $\cos (\omega_n\tau_0)\approx 1$ in the first term.]
Since this integral now converges, we may take $\tau_0\to 0$, assuming that 
$|\omega_n|\tau_0\ll 1$. We may then rescale the integral giving:
\begin{equation} \delta \GG = {-2\lambda^2\over \pi^2\tau_0}+
{2\lambda^2|\omega_n|\over \pi^2}\int_0^\infty {ds\over s}\sin s + O(\tau_0).\end{equation}
Evaluating the integral gives:
\begin{equation} \delta \GG = {-2\lambda^2\over \pi^2\tau_0}+
{2\lambda^2|\omega_n|\over 2\pi}+O(\tau_0).\end{equation}
Note that we again obtain a term which is even in $\omega_n$ but singular, 
involving an absolute value. Analytically continuing to real 
frequency, using Eq. (\ref{omegacon}) we obtain the zero temperature 
DC conductance from 
Eq. (\ref{CDC}):
\begin{equation} C={\lambda^2\over \pi}\to {2e^2\over h}\lambda^2,\end{equation}
the same result than is obtained from the Landauer formula. (Again 
a factor of $e^2/\hbar$ was inserted at the last step.) 

Alternatively, we could cut off the integral in Eq. (\ref{deltaG}), by 
a finite bandwidth, $2D$. Then the fermion propogator is modified to:
\begin{equation} <\psi^{\dagger \alpha}_i(\tau )\psi_{\beta j}(0)>\to \delta^\alpha_\beta 
\delta_{ij}{1-e^{-D|\tau |}\over 2\pi \tau},\end{equation}
giving:
\begin{equation} \delta G(\omega_n)=4\lambda^2\int_{-\infty}^\infty 
{d\tau e^{i\omega_n\tau}\left( 1-e^{-D|\tau |}\right)^2 \over (2\pi \tau )^2},\label{deltaG2}\end{equation}
where the integration region is now the entire real line. This can be expressed in 
terms of the exponential integral function, $E_i(x)$, giving a term of $O(D)$, 
the same universal term, $-\lambda^2|\omega_n|/\pi$, plus terms that 
vanish when $1/D\to 0$. This confirms that the term $\propto |\omega_n|$, which 
determines the conductance, is 
indeed universal. 

\section{Definitions of $T_K$}
The definition of $T_K$ used in this paper, and  defined 
in Eq. (\ref{Hint}), is the one first introduced by Nozi\`eres~\cite{Nozieres} 
in the original paper on local Fermi Liquid Theory for the Kondo model.  It 
corresponds to an arbitrary choice but does have the advantage that 
the coefficient of $T^2/T_K^2$ in the conductance 
has the relatively simple value of $\pi^2$.  [See Eq. (\ref{condfin}).]
Another popular definition is the one adopted by Wilson, which 
we refer to as $T_K^W$. This is fixed by the requirement that 
the impurity susceptibility, at $T\gg T_K$ have a particular 
form:
\begin{eqnarray}
\lefteqn{\chi_{imp}\to}&&\nonumber\\
&&{(g\mu_B)^2\over 4T} \bigg[1-{1\over \ln (T/T_K^W)}
-{(1/2)\ln \left[\ln (T/T_K^W)\right]\over \ln^2(T/T_K^W)}\nonumber\\
&&+O\left[\ln^2\left[\ln (T/T_K^W)\right]/\ln^3(T/T_K^W)\right] \bigg].
\end{eqnarray}
At low temperatures,
\be \chi_{imp}\to {(g\mu_B)^2w^2\over 4T_K^W},\ee
where the Wilson number, $w$ has the value:
\be w =e^{C+1/4}/\pi^{3/2}\approx .4128 \ee
and $C$ is Euler's constant ($\approx .577216$). 
These definitions of $T_K$ are related by:
\be T_K^W=(\pi w/4)T_K.\ee
Rather accurate results have been determined for the conductance 
at all temperatures using numerical renormalization group~\cite{Costi} 
and integrability.\cite{Konik}  These authors generally use 
a Kondo temperature, $T_K^C$, related to the other ones by:
\be T_K^C=T_K^W/w=(\pi /4)T_K,\ee
motivated, perhaps, by the fact that the zero temperature 
impurity susceptibility now takes the simple form 
$\chi_{imp}\to (g\mu_B)^2/(4T_K^C)$. 
Some experimental papers~\cite{GG,Wiel}
 on the Kondo effect in embedded quantum dots define 
a Kondo temperature by the condition that the conductance take 
half its zero temperature value at $T=T_K^e$:
\be C(T=T_K^e)/C(T=0)=1/2.\ee
The theoretical results in [\onlinecite{Konik}] indicate 
that $T_K^e\approx T_K^C$.
In [\onlinecite{AL}] we wrote the Fermi liquid interaction:
\be H_{int}=-\lambda \left (\psi_e^\dagger {\vec \sigma \over 2}\psi_e
\right)^2,\ee
(where $\lambda$ is not to be confused with the coefficient 
of the particle-hole breaking term in the current paper) 
and we normalized our fermions in an unconventional way 
so that $\psi_{AL}=\sqrt{2\pi}\psi$. Thus:
\be \lambda = {2\over 3\pi T_K}.\ee
Numerous other definitions of $T_K$ are also in use 
including, for example, the frequency scale 
at which Im${\cal T}(\omega ,T=0)$ is reduced by 1/2 from 
its maximum at zero frequency.

\acknowledgments
This research is supported by NSERC, CFI, SHARCNET and CIAR.
IA acknowledges interesting conversations with P. Simon.
\bibliography{flt}

\end{document}